%% file: 3dvector-arxiv-v2.tex
\documentclass[aps,11pt,onecolumn,nofootinbib,preprintnumbers, superscriptaddress]{revtex4-2}
\pdfoutput=1

\usepackage[height=8.85in,width=6.45in]{geometry}

\usepackage[utf8]{inputenc}
\usepackage[T1]{fontenc}
\usepackage{amsmath}
\usepackage{amssymb}
\usepackage{mathtools}
\usepackage{physics}
\usepackage{enumitem}
\allowdisplaybreaks

\usepackage{slashed}
\usepackage{braket}
\usepackage[usenames,dvipsnames,svgnames,table]{xcolor}
\usepackage[colorlinks,citecolor=DarkGreen,linkcolor=FireBrick,urlcolor=FireBrick,linktocpage,breaklinks=true]{hyperref}
\hypersetup{
    pdfstartview={FitH},    
    pdftitle={Strings from vectors},    
    pdfauthor={Chi-Ming Chang, Sean Colin-Ellerin, Cheng Peng, Mukund Rangamani},     
}
\usepackage{graphicx}
\usepackage{tikz}

\usetikzlibrary{decorations.markings}
\tikzset{line/.style={line width=0.25mm},
curve/.style={line,smooth,tension=1},
->-/.style={decoration={
  markings,
  mark=at position #1 with {\arrow[>=stealth]{>}}},postaction={decorate}},
-<-/.style={decoration={
  markings,
  mark=at position #1 with {\arrow[>=stealth]{<}}},postaction={decorate}},
}

\usepackage{times}
\usepackage[scaled]{couriers}

\usepackage{bm}

\usepackage{xcolor}
\usepackage{mdframed}

\renewenvironment{figure}[1][]{
  \begin{originalfigure}[#1]
    \begin{mdframed}[linecolor=black!0,backgroundcolor=black!0]
}{
    \end{mdframed}
  \end{originalfigure}
}

\usepackage{subcaption}

\definecolor{MM1}{rgb}{0.368417, 0.506779, 0.709798}
\definecolor{MM2}{rgb}{0.880722, 0.611041, 0.142051}
\definecolor{MM3}{rgb}{0.560181, 0.691569, 0.194885}

\newcommand{\sfp}{\mathfrak{p}}
\newcommand{\sfs}{\mathfrak{s}}
\newcommand{\sfpb}{\bar{\mathfrak{p}}}
\newcommand{\sfsb}{\bar{\mathfrak{s}}}

\begin{document}

\title{Disordered vector models: from higher spins to incipient strings}

\author{Chi-Ming Chang}
\email{cmchang@tsinghua.edu.cn}
\affiliation{Yau Mathematical Science Center (YMSC), Tsinghua University, Beijing, 100084, China}
\affiliation{Beijing Institute of Mathematical Sciences and Applications (BIMSA), Beijing, 101408, China}

\author{Sean Colin-Ellerin}
\email{scolinellerin@ucdavis.edu}
\affiliation{Center for Quantum Mathematics and Physics (QMAP),
Department of Physics \& Astronomy, University of California, Davis, CA 95616, USA}

\author{Cheng Peng}
\email{pengcheng@ucas.ac.cn}
\affiliation{Kavli Institute for Theoretical Sciences (KITS) and CAS Center for Excellence in Topological Quantum Computation, University of Chinese Academy of Sciences, Beijing 100190, China}

\author{Mukund Rangamani}
\email{mukund@physics.ucdavis.edu}
\affiliation{Center for Quantum Mathematics and Physics (QMAP),
Department of Physics \& Astronomy, University of California, Davis, CA 95616, USA}

\begin{abstract}
We present a one-parameter family of large $N$ disordered models, with and without supersymmetry, in three spacetime dimensions. They interpolate from the critical large $N$ vector model dual to a classical higher spin theory, toward  a theory with a classical string dual.  We analyze the spectrum and OPE data of the theories. While  the supersymmetric model is always well-behaved the non-supersymmetric model is unitary only over a small parameter range. We offer some speculations on the origin of strings from the higher spins. 
\end{abstract}

\maketitle

\section{Introduction}
\label{sec:intro}

Whereas the  planar  expansion of large $N$ gauge theories \cite{tHooft:1973alw}, suggestive of string perturbation theory, motivates the holographic AdS/CFT correspondence \cite{Maldacena:1997re}, large $N$ vector models, which capture criticality in a wide class of physical systems, eg., liquid-vapor, superfluid, and  Curie transition in ferromagnets \cite{Brezin:1972fb,Wilson:1973jj,Pelissetto:2000ek,Moshe:2003xn}, are dual to the  higher spin gravity in an anti-de Sitter spacetime   \cite{Klebanov:2002ja,Sezgin:2003pt,Giombi:2009wh}. Theories interpolating
between the two limits eg., $\mathcal{N}=4$ SYM \cite{Haggi-Mani:2000dxu,Gaberdiel:2021qbb} or  Chern-Simons matter theories \cite{Chang:2012kt}  are at best understood at the two extremes.

Models with intermediate behaviour, like the Sachdev-Ye-Kitaev (SYK) model  \cite{Sachdev:1992fk,Kitaev:2015aa,Maldacena:2016hyu,Kitaev:2017awl} and its cousins \cite{Fu:2016vas,Murugan:2017eto,Bulycheva:2018qcp,Peng:2018zap,Chang:2018sve,Chang:2021fmd,Berkooz:2021ehv}, characterized by melonic  diagrams dominating the large $N$ limit, offer new perspectives. Quantum mechanical examples ($d=1$) capture features of semiclassical gravity \cite{Maldacena:2016upp,Jensen:2016pah}, while $d\geq2$ examples  have classical   finite tension string duals \cite{Murugan:2017eto,Chang:2021fmd}, owing to the lack of sparsity in the spectrum and sub-maximal Lyapunov exponent.\footnote{It has been argued in \cite{Heemskerk:2009pn} that a necessary and sufficient criterion for a field theory to have a semiclassical gravity dual is that it have a sensible large $N$ limit and a sparse low-lying spectrum. One expects such theories to have maximal Lyapunov exponent \cite{Maldacena:2015waa}.}  We construct herein a one-parameter family of solvable three-dimensional (3d) theories (cf., \cite{Peng:2018zap} for  2d examples) where the higher spin symmetry gets Higgsed as we turn on the deformation. The higher spin states, however, remain in the spectrum suggesting an emergent string theory with finite tension.

We consider examples with two sets of fields transforming as vectors under ${\rm O}(N)$ and ${\rm O}(M)$, respectively (indexed by $i,j=1,\ldots, N$ and $a,b=1,\ldots,M$). We will discuss in parallel two sets of models:
\begin{itemize}[wide,left=0pt]
\item An $\mathcal{N}=2$ supersymmetric (susy) model with chiral superfields $\sfp^i$ and $\sfs^a$.
\item A bosonic (bos) model with fields $\phi^i$ and $\sigma^a$, obtained from the above, by retaining just the real parts of the bottom component of $\sfp^i$ and the top component of $\sfs^a$.
\end{itemize}

The dynamics of the two models is characterized by the (Euclidean) Lagrangian densities\footnote{We are employing Einstein summation convention for repeated indices and $y^\mu=x^\mu - i\, \theta\, \sigma^\mu \bar{\theta}$ is a chiral superspace coordinate. Our 3d $\mathcal{N}=2$ susy conventions are as in \cite{Chang:2021fmd}.} 
\begin{widetext}
\begin{equation}\label{eqn:action}
\begin{split}
\mathcal{L}_\text{susy} 
&= 
   -\int d^2 \theta \, d^2\bar{\theta}\; 
  \bigg( \bar{\sfp}_i(y^\dagger)\,\sfp^i(y) + \bar{\sfs}_a(y^\dagger)\,\sfs^a(y) \bigg)
  - \left[\int\, d^2\theta\, \frac{1}{2}\, g_{aij}\,\sfs^a(y) \, \sfp^i(y) \,\sfp^j(y)  + \text{c.c} \right] . \\
\mathcal{L}_\text{bos}
&=
 \frac{1}{2}\, \partial_\mu \phi_i \, \partial^\mu \phi^i+
 \frac{1}{2}\, g_{aij}\sigma^a \phi^i\phi^j-\frac{1}{4} (\sigma^a )^2  .
\end{split}
\end{equation}
\end{widetext}
The couplings $g_{aij}$ are  Gaussian random variables with zero mean and variance
\begin{equation}\label{eq:gGaussian}
\expval{ g_{aij}g_{bkl}}= \frac{2J}{N^2} \, \delta_{ab}\, \delta_{i(k}\delta_{l)j}\,, 
\qquad [J]_\text{classical} =1\,.
\end{equation}
The bosonic model has a positive semi-definite Hamiltonian; classically integrating out the auxiliary field $\sigma^a$ results in a vector model with a random quartic potential 
\begin{equation}\label{eq:quarticV}
V(\phi)=\frac{1}{4} \sum_{a=1}^M\left(\sum_{i,j=1}^Ng_{aij} \phi^i\phi^j\right)^2\,.
\end{equation}

When $M=1$ the random coupling $g_{1ij}$ can be absorbed by a ${\rm GL}(N,{\mathbb R})$ field redefinition reducing to the critical vector model or its ${\cal N}=2$ susy cousin \cite{Bobev:2015vsa,Bobev:2015jxa,Chester:2015qca,Chester:2015lej}. We will solve the models to leading order in the $1/N$ expansion while holding  \emph{'t Hooft coupling} $\lambda$ fixed:
\begin{equation}
N \to \infty \,, \qquad \lambda \equiv \frac{M}{N} \,, \quad \text{fixed}\,.
\end{equation}  
The $\lambda\to \infty$ limit is a variant of the bosonic 3d SYK model
with $q=4$  (${\rm bSYK}^{3d}_{q=4}$) \cite{Liu:2018jhs}. The ${\cal N}=2$ susy 3d SYK model \cite{Chang:2021fmd} is obtained for $\lambda = \frac{1}{2}$.

\section{The IR fixed points}
\label{sec:ircft}
 
The models can be solved analogously to the SYK model \cite{Kitaev:2015aa,Maldacena:2016hyu} by realizing the Schwinger-Dyson equations truncate.\footnote{While for the susy model one can use non-renormalization theorems to argue that the renormalization group flow does not induce operators that invalidate the melonic iteration, we are, strictly speaking, assuming such is also true in the bosonic model.} We illustrate the calculations for the bosonic model with the susy case generalizing straightforwardly by working with superfields, cf., \cite{Chang:2021fmd}. Some details are given in the Supplemental Material (Appendix~\ref{sec:details}).

\begin{figure}[h]
\centering
\begin{tikzpicture}[scale=0.4]
\draw [thick](-2,0)--(-0.6,0);
\draw [thick] (0.6,0) -- (2,0);
\draw [thick,lightgray, fill=lightgray] (0,0) circle (0.6cm);
\draw (0,0) node{$\scriptstyle{G}_\phi$};

\draw (2.5,0) node{$=$};

\draw [thick](3,0)--(6,0);
\draw (6.5,0) node{$+$};

\draw [thick](7,0)--(8,0);
\draw[thick,red] (8,0) to[out=60,in=210] (8.9,1);
\draw[thick,pink, fill=pink] (9.5,1) circle (0.6cm);
\draw[red] (9.5,1) node{$\scriptstyle{G}_\sigma$};
\draw[thick,red]  (10.1,1) to[out=-30,in=120] (11,0);

\draw[thick] (8,0) to[out=-60,in=150] (8.9,-1);
\draw[thick,lightgray, fill=lightgray] (9.5,-1) circle (0.6cm);
\draw (9.5,-1) node{$\scriptstyle{G}_\phi$};
\draw[thick]  (10.1,-1) to[out=30,in=-120] (11,0);

\draw [thick](11,0)--(11.9,0);
\draw[thick,lightgray, fill=lightgray] (12.5,0) circle (0.6cm);
\draw (12.5,0) node{$\scriptstyle{G}_\phi$};
\draw [thick](13.1,0)--(14,0);

\begin{scope}[shift={(0,-4)}]
\draw [thick,red](-2,0)--(-0.6,0);
\draw [thick,red] (0.6,0) -- (2,0);
\draw [thick,pink, fill=pink] (0,0) circle (0.6cm);
\draw [red](0,0) node{$\scriptstyle{G}_\sigma$};

\draw (2.5,0) node{$=$};

\draw [thick,red](3,0)--(6,0);
\draw (6.5,0) node{$+$};

\draw [thick,red](7,0)--(8,0);
\draw[thick] (8,0) to[out=60,in=210] (8.9,1);
\draw[thick,lightgray, fill=lightgray] (9.5,1) circle (0.6cm);
\draw (9.5,1) node{$\scriptstyle{G}_\phi$};
\draw[thick]  (10.1,1) to[out=-30,in=120] (11,0);

\draw[thick] (8,0) to[out=-60,in=150] (8.9,-1);
\draw[thick,lightgray, fill=lightgray] (9.5,-1) circle (0.6cm);
\draw (9.5,-1) node{$\scriptstyle{G}_\phi$};
\draw[thick]  (10.1,-1) to[out=30,in=-120] (11,0);

\draw [thick,red](11,0)--(11.9,0);
\draw[thick,pink, fill=pink] (12.5,0) circle (0.6cm);
\draw [red](12.5,0) node{$\scriptstyle{G}_\sigma$};
\draw [thick,red](13.1,0)--(14,0);
\end{scope}
\end{tikzpicture}
\caption{\small{Diagrammatic representation of the  Schwinger-Dyson equations.}}
\label{fig:SDprop}
\end{figure}
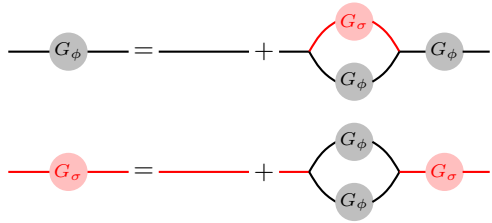

The two-point functions  $\expval{ \phi^i (x)\phi^j(0)} =\delta^{ij}G_\phi(x)$ and $\expval{ \sigma^a (x)\sigma^b (0)} =\delta^{ab}G_\sigma(x)$ are obtained by iterating melonic diagrams (Fig.~\ref{fig:SDprop}) leading to
\begin{equation}
\begin{aligned}
&
  G_\phi(p)= \frac{1}{p^2-\Sigma_\phi(-p)}\,,&& 
  G_\sigma(p)= -\frac{1}{\frac{1}{2}+\Sigma_\sigma(-p)} \,,
\\
&
  \Sigma_\phi(x)=\lambda \,J\, G_\phi(x)G_\sigma(x)\,,&&
  \Sigma_\sigma(x)= \frac{1}{2} J \,G_\phi(x)^2
\,.
\end{aligned}
\end{equation}
At scales below that set by $J$ we can ignore the bare propagators. Picking a conformal ansatz
\begin{equation}\label{eqn:conformal_2pt}
G_\phi(x) = \frac{b_\phi}{ |x|^{2\Delta_\phi}}\,,\qquad 
G_\sigma(x)= \frac{b_\sigma}{ |x|^{2\Delta_\sigma}}\,,
\end{equation}
we  solve for the scaling dimensions and one combination of the normalization coefficients.  We  find\footnote{In the susy model $\Delta_\sfp$ and $\Delta_\sfs$ refer to the conformal dimension of the chiral multiplet which is also that of the scalar field in the bottom component.}
\begin{equation}\label{eqn:dimension_phi}
\begin{aligned}
& \Delta_\sigma 
= 
  3 - 2\, \Delta_\phi\,, \qquad \Delta_\sfs =2-2\,\Delta_\sfp \,, \\ 
\lambda_\text{bos}
&= 
  \frac{\left(\Delta _{\phi }-2\right) \left(2 \Delta _{\phi }-3\right) \left(1+\sec \left(2
   \pi  \Delta _{\phi }\right)\right)}{4   \left(2 \Delta _{\phi } \left(4 \Delta
   _{\phi }-5\right)+3\right)} \,, \\
\lambda_\text{susy}
&= \frac{(\Delta_\sfp -1) \left(1+\sec\left(2\pi \Delta_\sfp \right)\right)}{2\,(2\, \Delta_\sfp-1)} \,.
\end{aligned}
\end{equation}
For a fixed 't Hooft coupling $\lambda$, the susy model has an unique solution satisfying the unitarity bound $\Delta_\sfp,\,\Delta_\sfs \ge \frac{1}{2}$, while the bosonic model has multiple solutions of the dimensions $\Delta_\phi$ and $\Delta_\sigma$. We focus on the branch continuously connected to the $\lambda=0$ theory.\footnote{The existence of the second branch of solutions signals potentially distinct degenerate vacua of \eqref{eqn:action}. Along 
this branch $\Delta_\phi \in (\frac{3}{4}, \frac{5}{4})$ naively contradicting the quartic interaction picture of \eqref{eq:quarticV}. One can motivate this by promoting $\sigma^a$ to a dynamical field and fine-tune  its mass term away as we flow down to attain the critical point. \label{fn:bosbranches} }

\begin{figure}[h]
\centering
\includegraphics[width=0.4\textwidth]{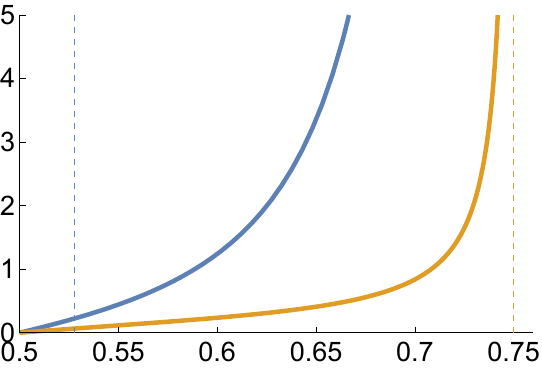}
 \setlength{\unitlength}{0.1\columnwidth}
\begin{picture}(0.3,0.4)(0,0)
\put(-4.3,2.5){\makebox(0,0){$\scriptstyle{\lambda}$}}
\put(0,0){\makebox(0,0){$\scriptstyle{\Delta}$}}
\put(-1.8,2.4){\makebox(0,0){$\color{MM1}{\Delta_\phi}$}}
\put(-0.7,2){\makebox(0,0){$\color{MM2}{\Delta_\sfp}$}}
\end{picture}
\caption{\small{Scaling dimensions $\Delta_\phi$ and $\Delta_\sfp$  as we vary $\lambda$. The bosonic model is unitary for $\Delta_\phi \in [0.5,0.52765)$ (region left of the dashed vertical line).  }}
\label{Fig:dimension}
\end{figure}
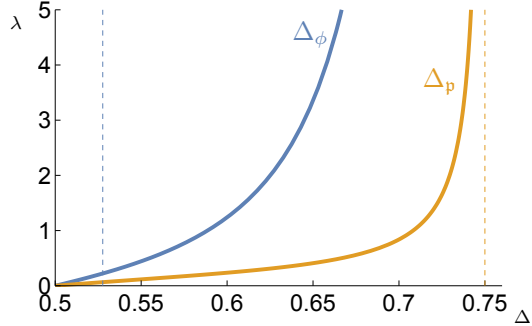

In Fig.~\ref{Fig:dimension} we plot the scaling dimension of the ${\rm O}(N)$ vectors in the two models.   Some salient features of interest are
\begin{itemize}[wide,left=0pt]
\item $\lambda=0$  in both cases corresponds to the the critical ${\rm O}(N)$ vector models with $\phi^i$ and $\sfp^i$ having free field dimensions while $\Delta_\sigma=2 $ and $\Delta_\sfs = 1$.
\item The bosonic model limits ${\rm bSYK}^{3d}_{q=4}$ as $\lambda \to \infty$ with $(\Delta_\phi , \Delta_\sigma) \to (\frac{3}{4}, \frac{3}{2})$. In the susy model we find 
$(\Delta_\sfp , \Delta_\sfs) \to (\frac{3}{4}, \frac{1}{2})$, whence $\sfs$ becomes a free field.
\item The intermediate value $\lambda = \frac{1}{2}$ gives $(\Delta_\sfp , \Delta_\sfs)= (\frac{2}{3}, \frac{2}{3})$, related to the fixed point of  \cite{Chang:2021fmd}. The bosonic theory is  related to ${\rm bSYK}^{3d}_{q=3}$ with $(\Delta_\phi , \Delta_\sigma)= (1, 1)$  but lies on a different branch of solutions.
\end{itemize}

\section{Single-trace operator spectrum}
\label{sec:spectrum}

An advantage of the disordered models is that one can obtain the spectrum of single trace operators and OPE coefficients. To do so we look at four-point functions, the connected contribution to which, denoted $\mathcal{F}$, is obtained by summing over the ladder diagrams and suitably diagonalizing the space of four-point correlators (see Supplemental Material \ref{sec:details}). We focus here  for simplicity on the singlet channel, which can be motivated by averaging over the external operators (as is common in the SYK literature). There  also are non-singlet channels from the tensor product of two vector representations (of $O(N)$ or $O(M))$ in the theory, which are qualitatively similar; cf., \cite{Chang:2021fmd}. We also note that there is no hierarchical separation between the singlets and the non-singlets in the large $N$ limit. 

Expanding in the (super)conformal partial wave basis one can write $\mathcal{F}$ in terms of a contour integral involving the (super)conformal  blocks, a spectral function $\rho(\Delta,\ell)$, and a ladder kernel $k(\Delta,\ell)$, as in various earlier explorations \cite{Maldacena:2016hyu,Murugan:2017eto}. We can schematically write (cf., Appendix~\ref{sec:details})
\begin{equation}\label{eq:Fcontour}
\mathcal{F} = \frac{1}{N} \, \sum_{\ell} \, \oint\, \frac{d\Delta}{2\pi i }\, \frac{\rho(\Delta,\ell)}{1-k(\Delta,\ell)} \, \mathcal{G}_{\Delta,\ell}\,.
\end{equation}  
The contour of integration for $\Delta$ is along the principal series line for (super)conformal  representations, i.e., along $\Delta = \frac{3}{2}+ i\, \mathbb{R}$ for the bosonic, and $\Delta = \frac{1}{2}+i\, \mathbb{R}$ for the susy model and closing toward  $\Delta \to +\infty$. This picks up the residues at the poles dictated by
\begin{equation}
k(\Delta,\ell)=1 \quad{\rm for}  \quad 
\begin{cases}
& \Re(\Delta_\text{bos})> \frac{3}{2} \,, \\
& \Re(\Delta_\text{susy})> \frac{1}{2} \,,
\end{cases}
\end{equation}
giving thence the spectrum of the single-trace operators. The residues at the poles are the squares of the OPE coefficients.   We discuss the two models in turn below.

\paragraph{Bosonic model:}  For the bosonic model we consider  correlators involving both $\phi^i$ and $\sigma^a$ and obtain the ladder kernel entering \eqref{eq:Fcontour}.  In the limit 
$\lambda \to \infty$ ($\Delta_\phi\to  \frac{3}{4}$), the ladder kernel at generic value\footnote{There are isolated special points where the limit \eqref{eqn:limit_to_q=4} does not hold, for example, at $\Delta=3$ and $\ell=0$.} of $\Delta$ coincides with the one of ${\rm bSYK}^{3d}_{q=4}$\footnote{Curiously, along the branch not connected to the free theory for $\Delta_\phi=1$ we encounter a simple relation to the kernel of ${\rm bSYK}^{3d}_{q=3}$: 
$\left[1-k(\Delta,\ell)\right]
=\left(1-k_{{\rm bSYK}^{3d}_{q=3}}(\Delta,\ell)\right)\left(1+ \frac{1}{2}k_{{\rm bSYK}^{3d}_{q=3}}(\Delta,\ell)\right) $.
}
\begin{equation}\label{eqn:limit_to_q=4}
\lim_{\lambda \to \infty} k(\Delta,\ell)=k_{{\rm bSYK}^{3d}_{q=4}}(\Delta,\ell)\,.
\end{equation}

\begin{figure}[h]
\centering
\subcaptionbox{Bosonic model.\label{Fig:HS_spectrum_N=0}}{
\hspace{-5mm}\includegraphics[width=0.5\textwidth]{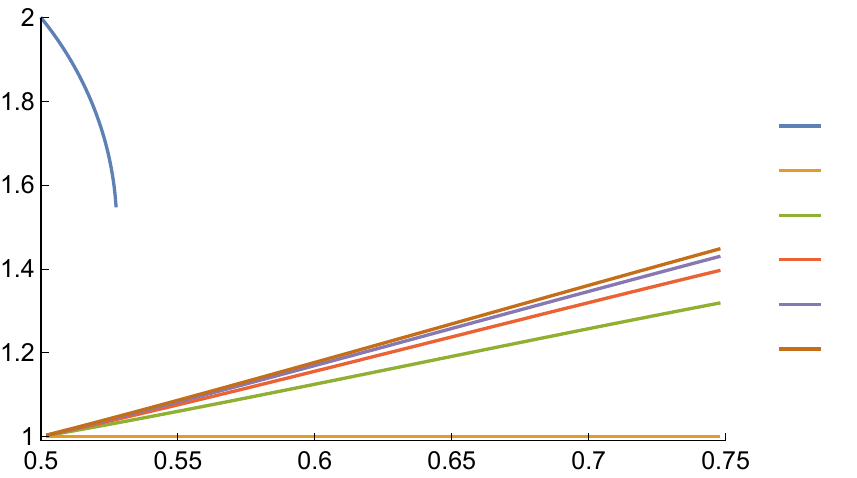}
\begin{picture}(0.3,0.4)(0,0)
\put(-200,130){\makebox(0,0){$\scriptstyle{\Delta-\ell}$}}
\put(-30,0){\makebox(0,0){$\scriptstyle{\Delta_\phi}$}}
\put(-18,98){\makebox(0,0){$\scriptstyle{\ell=0}$}}
\put(-18,86){\makebox(0,0){$\scriptstyle{\ell=2}$}}
\put(-18,75){\makebox(0,0){$\scriptstyle{\ell=4}$}}
\put(-18,63){\makebox(0,0){$\scriptstyle{\ell=6}$}}
\put(-18,52){\makebox(0,0){$\scriptstyle{\ell=8}$}}
\put(-16,40){\makebox(0,0){$\scriptstyle{\ell=10}$}}
\end{picture}}
\subcaptionbox{Susy model.\label{Fig:HS_spectrum_N=2}}{
\includegraphics[width=0.5\textwidth]{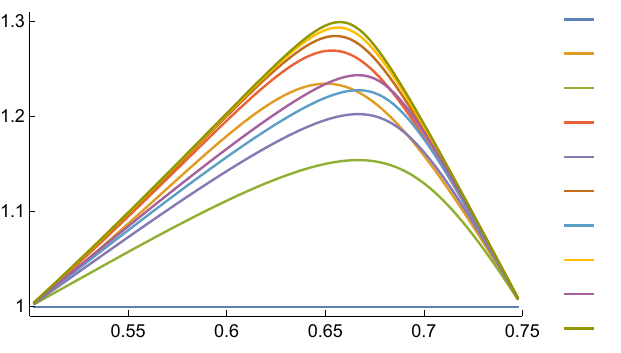}
\begin{picture}(0.3,0.4)(0,0)
\put(-200,130){\makebox(0,0){$\scriptstyle{\Delta-\ell}$}}
\put(-30,0){\makebox(0,0){$\scriptstyle{\Delta_\sfp}$}}
\put(-16,123){\makebox(0,0){$\scriptstyle{\ell=0,1}$}}
\put(-18,111){\makebox(0,0){$\scriptstyle{\ell=2}$}}
\put(-18,99){\makebox(0,0){$\scriptstyle{\ell=3}$}}
\put(-18,87){\makebox(0,0){$\scriptstyle{\ell=4}$}}
\put(-18,75){\makebox(0,0){$\scriptstyle{\ell=5}$}}
\put(-18,63){\makebox(0,0){$\scriptstyle{\ell=6}$}}
\put(-18,51){\makebox(0,0){$\scriptstyle{\ell=7}$}}
\put(-18,39){\makebox(0,0){$\scriptstyle{\ell=8}$}}
\put(-18,27){\makebox(0,0){$\scriptstyle{\ell=9}$}}
\put(-16,14){\makebox(0,0){$\scriptstyle{\ell=10}$}}
\end{picture}}
\caption{The dimensions of the leading twist operators. The spin-0 spectrum for the bosonic model is complex for $\Delta_\phi>0.528$ ($\lambda >0.222$).}
\label{Fig:HSspectrum}
\end{figure}
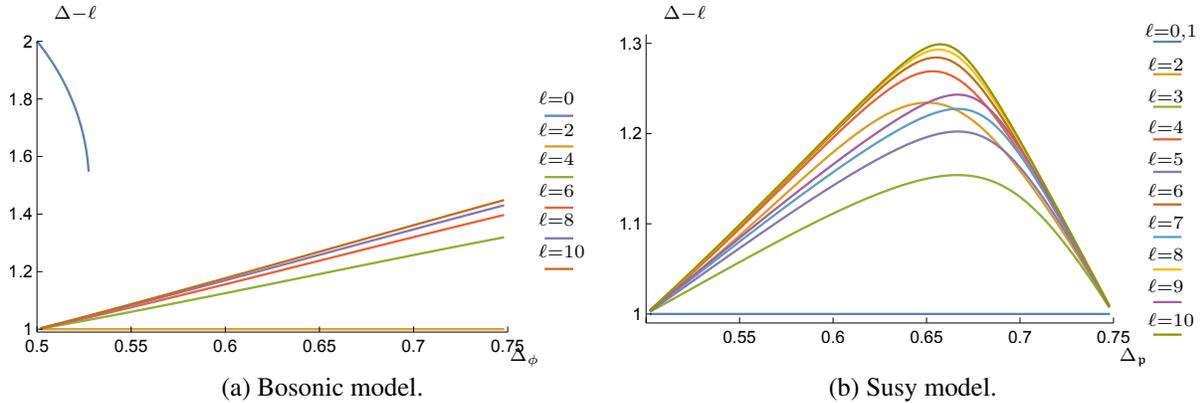

The spectrum is  organized into Regge trajectories,
\begin{equation}\label{eqn:spectrum}
\Delta=2\Delta_\alpha+\ell +2 n +\gamma_\alpha(\ell,n) \,, \quad \alpha\in\{\phi,\sigma\}\,,
\end{equation}
for $\ell\in 2{\mathbb Z}_{\ge 0}$ and $n\in {\mathbb Z}_{\ge 0}$. The operators on the  leading Regge trajectory (leading twist) have twist $\Delta-\ell$ behaving as depicted in Fig.~\ref{Fig:HS_spectrum_N=0}. The $\ell=0$ trajectory terminates at $\Delta_\phi=0.52765$ because the conformal  dimension of the operator becomes complex on the principal series $\Delta= \frac{3}{2} +i\nu$ when $\Delta_\phi>0.52765$ ($\lambda \gtrsim  0.222)$. This signals that the model becomes non-unitary beyond this point. Such behaviour was also observed in the bosonic SYK model  and tensor models in \cite{Giombi:2017dtl} and is consistent with the limiting behaviour noted in \eqref{eqn:limit_to_q=4}.\footnote{A similar feature is seen in bifundamental multiscalar models, see for example  \cite{Osborn:2017ucf} (and also  \cite{Kapoor:2021lrr} which we received as this article was in preparation).} The $\ell=2$ line with the constant unit twist corresponds to the stress tensor. The twists of the higher spin operators ($\ell>2$) increase along with $\Delta_\phi$.\footnote{The trajectories are smooth through $\Delta_\phi = \frac{3}{4}$ attaining maxima at $\Delta_\phi=1$, and then decreasing back to one at $\Delta_\phi = \frac{5}{4}$.} 

As $\lambda\to0$  the spectrum of the leading twist operators approaches that of the critical ${\rm O}(N)$ model as 
\begin{equation}
\Delta \to 
\begin{cases}
  2 &
    {\rm for}\;\ \ell=0\,, \\
  \ell+1+ \frac{16}{3\pi^2} \frac{\ell-2 }{3+2(\ell-2)} \, \lambda 
   &
    {\rm for}\;\ \ell=2,\,4,\,\cdots\,.
\end{cases}
\end{equation}
In addition,  besides the double twist operators with $\Delta\to\ell+4+2n$ in the $\sigma\sigma\to\sigma\sigma$ channel, operators in the subleading and higher Regge trajectories decouple from the spectrum, as their OPE coefficients approach zero, verifying indeed that as  $\lambda\to0$ we revert to the  critical ${\rm O}(N)$ model.

In the large spin and large twist limits, the anomalous dimensions $\gamma_\phi$ and $\gamma_\sigma$ scale as
\begin{equation}
\lim_{\ell \gg 1}\gamma_{\phi,\sigma}(\ell,n) \sim \frac{1}{ \ell^{2\Delta_\phi}}\,, 
\quad 
\lim_{n \gg 1}\gamma_{\phi,\sigma}(\ell,n) \sim \frac{1}{n^{4\Delta_\phi}}\,,
\end{equation}
consistent with the large spin analytic bootstrap \cite{Fitzpatrick:2012yx,Komargodski:2012ek}.  The central charge of the theory can be obtained  from the $\phi^i\phi^i\, T^{\mu\nu}$ OPE coefficient. We find:  
\begin{equation}\label{eq:CTbos0}
C_T \to 
N\left( \frac{3}{2}  -\frac{20}{3\pi^2} \, \lambda + \cdots\right) \ \text{as} \ \lambda \to 0\,,
\end{equation}  
as expected for a system of $N$ free bosons. 

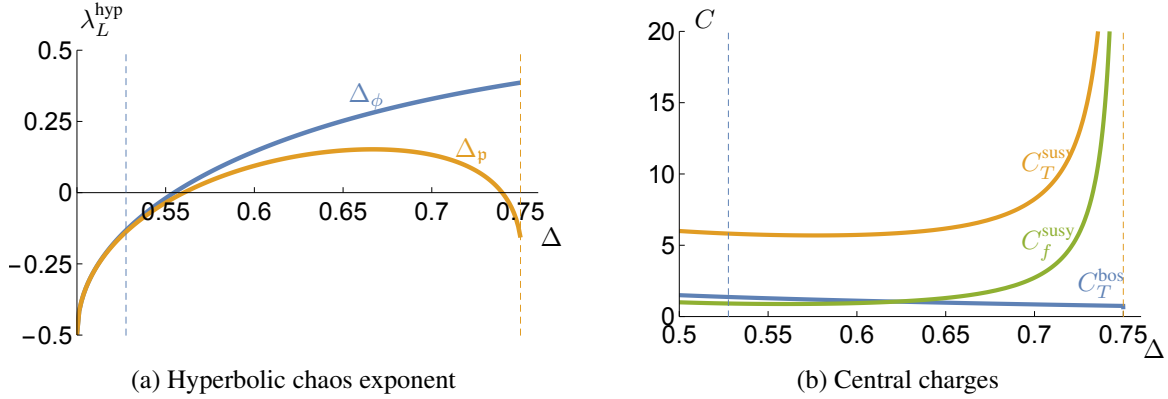
\begin{figure}[h]
\centering
\subcaptionbox{\small{Hyperbolic chaos exponent}\label{Fig:ChaosExponent}}{
\hspace*{-.45cm}\includegraphics[width=0.45625\textwidth]{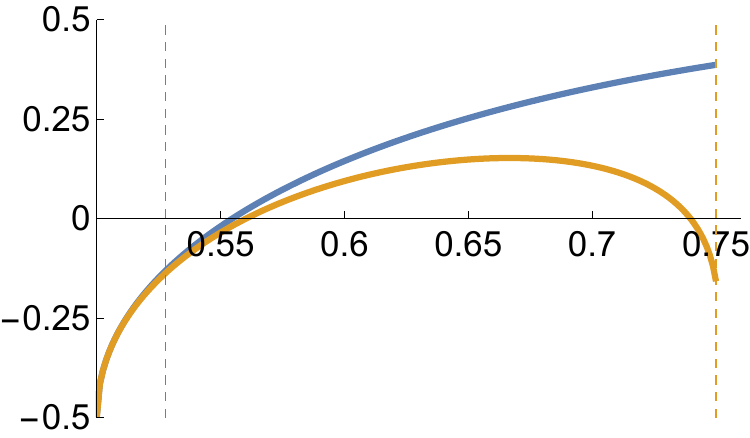} 
\begin{picture}(0.3,0.4)(0,0)
\put(-170,125){\makebox(0,0){$\lambda_L^\text{hyp}$}}
\put(0,42.5){\makebox(0,0){$\Delta$}}
\put(-70,95.5){\makebox(0,0){$\color{MM1}{\Delta_\phi}$}}
\put(-30,75){\makebox(0,0){$\color{MM2}{\Delta_\sfp}$}}
\end{picture}}
\hspace{1cm}
\subcaptionbox{\small{Central charges}\label{Fig:ccj}}{
\includegraphics[width=0.425\textwidth]{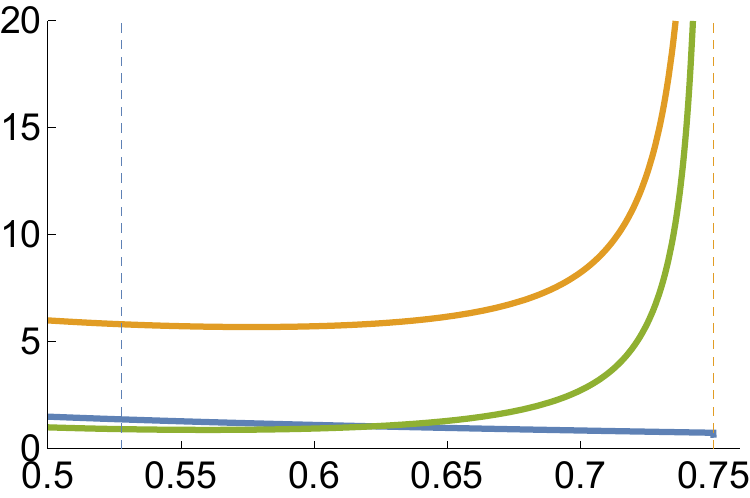} 
\begin{picture}(0.3,0.4)(0,0)
\put(-170,125){\makebox(0,0){$C$}}
\put(0,0){\makebox(0,0){$\Delta$}}
\put(-20,25){\makebox(0,0){$\color{MM1}{C_T^\text{bos}}$}}
\put(-40,70){\makebox(0,0){$\color{MM2}{C_T^\text{susy}}$}}
\put(-40,40){\makebox(0,0){$\color{MM3}{C_f^\text{susy}}$}}
\end{picture}}
\caption{\small{Hyperbolic chaos exponent $\lambda_L^\text{hyp}$ and central charges $C_{T}$ and $C_f$ for the two models}.}
\label{fig:cchypchaos}
\end{figure}

Beside the spectral information, the Lyapunov exponent $\lambda_L^\text{hyp}$ of the out-of-time-order four-point function in hyperbolic space is also encoded in $k(\Delta,\ell)$ \cite{Murugan:2017eto},
\begin{equation}\label{eq:lamhyp}
\lambda_L^\text{hyp} = \ell_* -1 \,,\qquad 
k\left( \frac{3}{2} ,\ell_*\right)=1\,.
\end{equation}
The behaviour of $\lambda_L^\text{hyp} $ is shown in Fig.~\ref{Fig:ChaosExponent}. At the two extreme ends $\Delta_\phi \to  \{\frac{1}{2}, \frac{3}{4}\}$ we find $\lambda^{\rm hyp}_L$ attains the value in critical ${\rm O}(N)$ model and ${\rm bSYK}^{3d}_{q=4}$, respectively.\footnote{We also recover the result for ${\rm bSYK}^{3d}_{q=3}$ at $\Delta_\phi =1$.}

\paragraph{Susy model:} The analysis of the susy model is similar though we work directly with superconformal blocks as in \cite{Bobev:2015jxa,Chang:2021fmd}. The results for the leading twist spectrum and hyperbolic chaos exponent are plotted in Figs.~\ref{Fig:HS_spectrum_N=2} and \ref{Fig:ChaosExponent}, respectively. The single trace spectrum is again organized into two Regge trajectories
\begin{equation}\label{eq:susyDeltagam}
\Delta = 2\,\Delta_\mathfrak{a} + \ell + 2n+ \gamma_\mathfrak{a}(\ell,n) \,, \quad 
\mathfrak{a} \in \{\sfp, \sfs\}
\end{equation}
with $\ell, n\in \mathbb{Z}_{\geq 0}$, cf., \eqref{eq:susyRanomdim}.

In limiting case $\Delta_\sfp\to  \frac{1}{2}$ we recover the ${\cal N}=2$ susy ${\rm O}(N)$ model \cite{Bobev:2015vsa,Bobev:2015jxa,Chester:2015qca,Chester:2015lej} with  $\lambda(\Delta_\sfp)$. The spectrum simplifies: in  the $\sfp \bar{\sfp} \to\sfp \bar{\sfp} $ channel we find a tower of higher spin currents at leading twist
\begin{equation}
\Delta \to  
  \ell + 1  + \frac{8}{\pi^2} \frac{2\,\ell-1+(-1)^\ell}{2\,\ell+1} \,\lambda .
\end{equation}  
The higher twist operators in this channel  decouple in the limit and only operators with $\Delta \to\ell+2 + 2n$ from the $\sfs \bar{\sfs} \to \sfs \bar{\sfs}$  channel survive. 

 At the other end, as $\Delta_\sfp \to \frac{2}{3}$ we encounter the  ${\cal N}=2$ susy 3d SYK model studied in \cite{Chang:2021fmd}.  The two theories have identical $\lambda^{\rm hyp}_L=0.15207$ with spectrum of the latter model being contained in ours. We however have two flavors of fields and thus also have  a $U(1)_f$ flavor symmetry wherein $q(\sfs) = -2\,q(\sfp)$ in addition to a $U(1)_R$ $R$-symmetry. The current multiplets are the $\ell=0,1$ trajectories in Fig.~\ref{Fig:HS_spectrum_N=2}. Computing the OPE coefficients of the current multiplets we find the central charges for the models consistent with results obtained using supersymmetric localization, cf., \eqref{eq:Csusy}.

\section{Discussion}
\label{sec:discuss}

We have at hand a one-parameter family of disordered models smoothly interpolating from the large $N$ critical ${\rm O}(N)$ vector model, breaking the higher spin symmetry as $\lambda >0$. The bosonic model is unitary for a small window  $\lambda \in[0,0.222)$, but the susy model is sensible for $\lambda\in \mathbb{R}_{\geq 0}$. 

For $\lambda >0$ the higher spin operators pick up non-vanishing anomalous dimensions, cf., Fig.~\ref{Fig:HSspectrum} leading one to expect a classical string dual description resulting from this Higgsing. In conventional AdS/CFT examples, the free field limit has been analyzed in several works \cite{Haggi-Mani:2000dxu,Mikhailov:2002bp,Gopakumar:2003ns,Chang:2012kt} with recent constructions of the worldsheet string description \cite{Eberhardt:2018ouy,Gaberdiel:2021jrv,Gaberdiel:2021qbb} but it is as yet unclear how to connect them to the supergravity description at strong coupling. While we do not yet have an explicit dual, the tractability of the models and the $\lambda=0$ limit being dual to higher spin AdS gravity offers tantalizing possibilities.

The higher spin states are always in the spectrum, so one expects a dual with a finite string tension. Moreover, their anomalous dimensions exhibit a power-law behaviour seen in analytic bootstrap \cite{Fitzpatrick:2012yx,Komargodski:2012ek} and not the logarithmic growth expected from semiclassical strings \cite{Gubser:2002tv}. As explained in \cite{Alday:2015ota} this may be attributed to the fact that vector models have operators with twists close to the unitarity bound.  

So how may we expect strings to emerge? A speculation we can offer is the following: at the higher spin limit the bulk degrees of freedom are the bilocal collective fields \cite{Das:2003vw,Aharony:2020omh}, the two-point functions $G(x_1,x_2)$. Let us visualize these bilocal objects to be one-dimensional with end points given by the two operators; we simply have a free Fock space of these collective fields. However, as we turn on $\lambda$ we should anticipate some linking between different bilocals, leading to a two-dimensional structure, an incipient worldsheet. The glue binding these worldsheets is not as strong as in planar gauge theories so we don't quite make it to the supergravity point. It remains to be seen how to flesh out these ideas, but having analytic control over the field theory is a promising starting point for  a perturbative analysis for small $\lambda $. 

We have focused on the IR fixed point vacuum,  but real-time thermal dynamics, be it retarded response, or out-of-time-order (OTO) observables, should give us clues about the nature of stringy black holes duals, through connections to quasinormal modes and Lyapunov exponents. A promising avenue would be to understand the mean-field description of OTO correlators \cite{Gu:2021xaj} to glean clues about the stringy dual. 

\begin{acknowledgments}

It is a pleasure to thank Micha Berkooz and Adar Sharon for discussions on disordered systems, and Xinan Zhou for discussions on anomalous dimensions.
CC is partly supported by National Key R\&D Program of China (NO. 2020YFA0713000).
SCE was  supported by U.S.\ Department of Energy grant DE-SC0019480  under the HEP-QIS QuantISED program and funds from the University of California. CP  is supported by the Fundamental Research Funds for the Central Universities, by funds from the University of Chinese Academy of Science, and NSFC NO. 12175237. MR  was supported by  U.S. Department of Energy grant DE-SC0009999 and by funds from the University of California.  
\end{acknowledgments}


\include{3dvector-refs}


\newpage
\appendix

\begin{widetext}

\section{Computational details} 
\label{sec:details}

We provide some additional details of our computations and outline some salient results described in the main text in this supplementary material. Our discussion will be brief; the reader can find additional details spelt out in the literature, cf., \cite{Murugan:2017eto,Liu:2018jhs,Chang:2021fmd}. 

\subsection{2-point functions} 

As noted around \eqref{eqn:conformal_2pt}  the iteration of  the melonic diagrams in Fig.~\ref{fig:SDprop} not only determines the conformal dimensions, but also a single combination of the normalization coefficients. We find:
\begin{equation}\label{eq:bbos}
\begin{split}
\mathfrak{b}_\text{bos}
& \equiv
   J\,b_\phi^2\, b_\sigma 
  = 
  \frac{16}{\pi^3}\, \left(\Delta _{\phi }-1\right) \left( \Delta _{\phi }-\frac{1}{2}\right) \left(\Delta
   _{\phi }-\frac{3}{4}\right) \cot \left(2 \pi  \Delta _{\phi }\right)  \,.
\end{split}
\end{equation}  
The calculation for the susy model parallels that of the bosonic model. We work with the superfields, noting that in terms of the  chiral coordinate of $\mathcal{N}=2$ 
superspace $y^\mu = x^\mu - i\, \theta\, \sigma^\mu \bar{\theta}$, the super-translation invariant is 
$z_{12}^\mu = y_1^\mu - y_2^\mu+2i\,\bar{\theta}_1 \sigma^\mu \theta_2$. Using the ansatz
\begin{equation}\label{eq:G2ptsusy}
G_\sfp(z) = \frac{b_\sfp}{ |z|^{2\Delta_\sfp}}\,,\qquad 
G_\sfs(z)= \frac{b_\sfs}{ |z|^{2 \Delta{_\sfs}} }\,,
\end{equation}  
we obtain the solution given in \eqref{eqn:dimension_phi} along with 
\begin{equation}\label{eq:bsusy}
\mathfrak{b}_\text{susy}
\equiv
  J\,b_\sfp^2\, b_\sfs
  = \frac{1}{2\pi^3} \, \left(2\, \Delta_\sfp-1\right) \, \cot(2\pi\, \Delta_\sfp)\,.
\end{equation}  
%

\subsection{4-point functions:} 

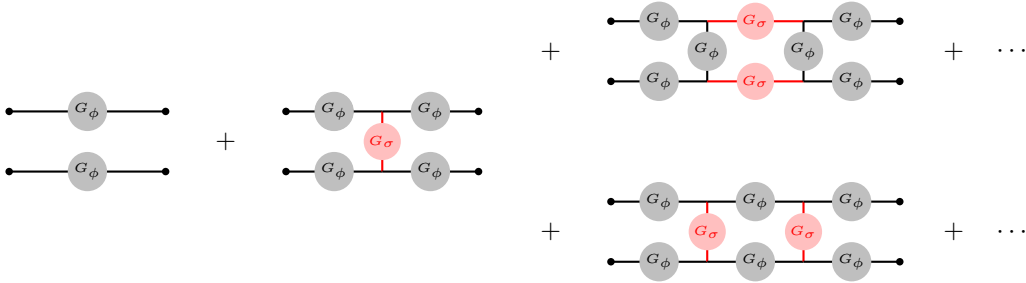
\begin{figure}[htbp]
\centering
\begin{tikzpicture}[scale=0.8]

\draw[thick, shift={(-2.6,0.5)}] (0,0) to (1,0) (1.3,0)node[circle,fill=lightgray,inner sep=.04cm]{$\scriptscriptstyle{G}_\phi$} (1.6,0) to (2.6,0) (0,0) node[circle,fill,inner sep=1pt]{} (2.6,0) node[circle,fill,inner sep=1pt]{};
\draw[thick, shift={(-2.6,-0.5)}] (0,0) to (1,0) (1.3,0)node[circle,fill=lightgray,inner sep=.04cm]{$\scriptscriptstyle{G}_\phi$} (1.6,0) to (2.6,0)
(0,0) node[circle,fill,inner sep=1pt]{} (2.6,0) node[circle,fill,inner sep=1pt]{} ;

\draw (1,0) node{$+$};

\draw [thick,gray, fill=gray, shift={(2,.5)}] (.8,0) circle (.3cm) ;
\draw[thick, shift={(2,.5)}] (0,0) to (.5,0) (.8,0)node[circle,fill=lightgray,inner sep=.04cm]{$\scriptscriptstyle{G}_\phi$} (1.1,0) to (1.6,0) (0,0) node[circle,fill,inner sep=1pt]{};
\draw[thick, shift={(2,-.5)}] (0,0) to (.5,0) (.8,0)node[circle,fill=lightgray,inner sep=.04cm]{$\scriptscriptstyle{G}_\phi$} (1.1,0) to (1.6,0) (0,0) node[circle,fill,inner sep=1pt]{};

\draw[thick,red,shift={(3.6,0)}] 
(0,-.5) to (0,-.3)
(0,.3) to  (0,.5)
(0,0)node[circle,fill=pink,inner sep=.04cm]{$\scriptscriptstyle{G}_\sigma$};

\draw[thick, shift={(3.6,.5)}] (0,0) to (.5,0) (.8,0)node[circle,fill=lightgray,inner sep=.04cm]{$\scriptscriptstyle{G}_\phi$} (1.1,0) to (1.6,0) (1.6,0) node[circle,fill,inner sep=1pt]{} ;
\draw[thick, shift={(3.6,-.5)}] (0,0)  to (.5,0)  (.8,0)node[circle,fill=lightgray,inner sep=.04cm]{$\scriptscriptstyle{G}_\phi$} (1.1,0) to (1.6,0) (1.6,0) node[circle,fill,inner sep=1pt]{} ;

\begin{scope}[shift={(9,1.5)}]

\draw (-2.6,0) node{$+$};

\draw[thick, shift={(-1.6,.5)}] (0,0) to (.5,0) (.8,0)node[circle,fill=lightgray,inner sep=.04cm]{$\scriptscriptstyle{G}_\phi$} (1.1,0) to (1.6,0) (0,0) node[circle,fill,inner sep=1pt]{};
\draw[thick, shift={(-1.6,-.5)}] (0,0) to (.5,0) (.8,0)node[circle,fill=lightgray,inner sep=.04cm]{$\scriptscriptstyle{G}_\phi$} (1.1,0) to (1.6,0) (0,0) node[circle,fill,inner sep=1pt]{};

\draw[thick,shift={(0,0)}] 
(0,-.5) to (0,-.3)
(0,0)node[circle,fill=lightgray,inner sep=.04cm]{$\scriptscriptstyle{G}_\phi$}
(0,.3) to  (0,.5);

\draw[thick,red, shift={(0,.5)}] (0,0) to (.5,0) (.8,0)node[circle,fill=pink,inner sep=.04cm]{$\scriptscriptstyle{G}_\sigma$} (1.1,0) to (1.6,0) ;
\draw[thick,red, shift={(0,-.5)}] (0,0)  to (.5,0)  (.8,0)node[circle,fill=pink,inner sep=.04cm]{$\scriptscriptstyle{G}_\sigma$} (1.1,0) to (1.6,0) (1.6,0) ;

\draw[thick,shift={(1.6,0)}] 
(0,-.5) to (0,-.3)
(0,0)node[circle,fill=lightgray,inner sep=.04cm]{$\scriptscriptstyle{G}_\phi$}
(0,.3) to  (0,.5);

\draw[thick, shift={(1.6,.5)}] (0,0) to (.5,0) (.8,0)node[circle,fill=lightgray,inner sep=.04cm]{$\scriptscriptstyle{G}_\phi$} (1.1,0) to (1.6,0) (1.6,0) node[circle,fill,inner sep=1pt]{};
\draw[thick, shift={(1.6,-.5)}] (0,0)  to (.5,0)  (.8,0)node[circle,fill=lightgray,inner sep=.04cm]{$\scriptscriptstyle{G}_\phi$} (1.1,0) to (1.6,0) (1.6,0) (1.6,0) node[circle,fill,inner sep=1pt]{};

\draw (4.1,0) node{$+$};
\draw (5.1,0) node{$\cdots$};

\end{scope}

\begin{scope}[shift={(9,-1.5)}]

\draw (-2.6,0) node{$+$};

\draw[thick, shift={(-1.6,.5)}] (0,0) to (.5,0) (.8,0)node[circle,fill=lightgray,inner sep=.04cm]{$\scriptscriptstyle{G}_\phi$} (1.1,0) to (1.6,0) (0,0) node[circle,fill,inner sep=1pt]{};
\draw[thick, shift={(-1.6,-.5)}] (0,0) to (.5,0) (.8,0)node[circle,fill=lightgray,inner sep=.04cm]{$\scriptscriptstyle{G}_\phi$} (1.1,0) to (1.6,0) (0,0) node[circle,fill,inner sep=1pt]{};

\draw[thick,red,shift={(0,0)}] 
(0,-.5) to (0,-.3)
(0,0)node[circle,fill=pink,inner sep=.04cm]{$\scriptscriptstyle{G}_\sigma$}
(0,.3) to  (0,.5);

\draw[thick, shift={(0,.5)}] (0,0) to (.5,0) (.8,0)node[circle,fill=lightgray,inner sep=.04cm]{$\scriptscriptstyle{G}_\phi$} (1.1,0) to (1.6,0) ;
\draw[thick, shift={(0,-.5)}] (0,0)  to (.5,0)  (.8,0)node[circle,fill=lightgray,inner sep=.04cm]{$\scriptscriptstyle{G}_\phi$} (1.1,0) to (1.6,0) (1.6,0) ;

\draw[thick,red,shift={(1.6,0)}] 
(0,-.5) to (0,-.3)
(0,0)node[circle,fill=pink,inner sep=.04cm]{$\scriptscriptstyle{G}_\sigma$}
(0,.3) to  (0,.5);

\draw[thick, shift={(1.6,.5)}] (0,0) to (.5,0) (.8,0)node[circle,fill=lightgray,inner sep=.04cm]{$\scriptscriptstyle{G}_\phi$} (1.1,0) to (1.6,0) (1.6,0) node[circle,fill,inner sep=1pt]{};
\draw[thick, shift={(1.6,-.5)}] (0,0)  to (.5,0)  (.8,0)node[circle,fill=lightgray,inner sep=.04cm]{$\scriptscriptstyle{G}_\phi$} (1.1,0) to (1.6,0) (1.6,0) (1.6,0) node[circle,fill,inner sep=1pt]{};

\draw (4.1,0) node{$+$};
\draw (5.1,0) node{$\cdots$};

\end{scope}

\end{tikzpicture}
\caption{\small{ Ladder iteration to obtain  $\expval{\phi(x_1) \, \phi(x_2)\, \phi(x_3)\, \phi(x_4)}$ in the bosonic model. }}
\label{fig:4ptfnladder}
\end{figure}

The computation of the 4-point functions for these disordered models is made feasible by iterating ladder diagrams, a special case of which is illustrated for the bosonic model 4-point functions of $\phi^i$ in Fig.~\ref{fig:4ptfnladder}. Focusing on the bosonic model, the set of averaged and normalized four-point functions at hand, eg., 
\begin{equation}\label{eq:4ptconnav}
 \frac{1}{N^2\, G_\phi(x_{12})\, G_\phi(x_{34})  } \sum_{i,j=1}^N\expval{\phi^i(x_1) \phi^i(x_2)\phi^j(x_3)\phi^j(x_4)}  = 1 + \frac{1}{N}\, \mathcal{F}^{\phi\phi\phi\phi}(\{x_i\})  \,, 
\end{equation} 
and its cousins involving $\sigma^a$ (and similarly for the susy model) can be assembled into a matrix:
\begin{equation}\label{eq:4ptmatrix}
\bm{\mathcal{F}}_\text{bos} = 
\begin{pmatrix}
b_\phi^2\mathcal{F}^{\phi\phi\phi\phi} & b_\phi b_\sigma\mathcal{F}^{\phi\phi\sigma\sigma} \\
b_\phi b_\sigma\mathcal{F}^{\sigma\sigma\phi\phi} &  b_\sigma^2\mathcal{F}^{\sigma\sigma\sigma\sigma}
\end{pmatrix}
\,, \qquad 
\bm{\mathcal{F}}_\text{susy} = 
\begin{pmatrix}
b_\sfp^2 \mathcal{F}^{\sfpb\sfp\sfp\sfpb} &  b_\sfp b_\sfs \mathcal{F}^{\sfpb\sfp\sfs\sfsb} \\
b_\sfp b_\sfs\mathcal{F}^{\sfsb\sfs\sfp\sfpb} & b_\sfs^2\mathcal{F}^{\sfsb\sfs\sfs\sfsb}.
\end{pmatrix}
\end{equation}  
The iteration depicted in Fig.~\ref{fig:4ptfnladder} implies that these correlators are obtained by the action of the ladder kernel operator on the disconnected 4-point function:
\begin{equation}\label{eq:4ptladders}
\begin{split}
\bm{\mathcal{F}}_\text{bos} &= \sum_{n=0}^\infty\, 
  \begin{pmatrix}
  K^{\phi\sigma\phi}  & K^{\phi\phi\phi} \\
  K^{\sigma\phi\sigma}  & 0 \\
  \end{pmatrix}^{*n} * 
  \begin{pmatrix}
  b_\phi^2 \mathcal{F}^{\phi\phi\phi\phi}_0 & 0 \\
  0 & b_\sigma^2\mathcal{F}^{\sigma\sigma\sigma\sigma}_0
  \end{pmatrix}, \\
\bm{\mathcal{F}}_\text{susy} &= \sum_{n=0}^\infty\,   \begin{pmatrix}
  K^{\sfp\sfs\sfp}  & K^{\sfp\sfp\sfp} \\
  K^{\sfs\sfp\sfs}  & 0 \\
  \end{pmatrix}^{*n}   * 
  \begin{pmatrix}
 b_\sfp^2 \mathcal{F}^{\sfpb\sfp\sfp\sfpb}_0 & 0 \\
  0 & b_\sfs^2 \mathcal{F}^{\sfsb\sfs\sfs\sfsb}_0
 \end{pmatrix} .
  \end{split}
  \end{equation}  
The product notation `$*$' subsumes both matrix multiplication and the action of the kernel operator on the disconnected diagram which involves integration over the intermediate positions. We note in passing that one can consider more general correlators
$\expval{\phi^i(x_1) \phi^j(x_2)\phi^k(x_3)\phi^l(x_4)}$ which would give us access to the non-singlet channel OPE. These were analyzed in detail for the single field model in \cite{Chang:2021fmd} and we expect the behaviour here to be qualitatively similar. 

The eigenvalues of the individual kernel operators, denoted $k^{\phi\phi\phi}$, $k^{\sfp\sfs\sfp}$ etc., are obtained by considering their action on (super)conformal three-point functions. Further, expanding the disconnected four-point functions in terms of (super)conformal partial waves we arrive at our final result:
\begin{equation}\label{eq:4ptfinal}
\begin{split}
& 
\bm{\mathcal{F}}_\text{bos}
= 
  \sum_{\ell=0}^{\infty}
    \int_{0}^{\infty}
      \frac{ds}{(1-k_\text{bos}^+)(1-k_\text{bos}^-)} \, 
      \begin{pmatrix} 
       b_\phi^2 \frac{\langle\mathcal{F}_{0}^{\phi\phi\phi\phi},\Theta_{\Delta,\ell}\rangle}{\langle\Theta_{\Delta,\ell},\Theta_{\Delta,\ell}\rangle} \, \Theta_{\Delta,\ell} 
        & 
          -\frac{k_\text{bos}^- \,k_\text{bos}^+}{k^{\sigma\phi\sigma}}b_\sigma^2\frac{\langle\mathcal{F}_{0}^{\sigma\sigma\sigma\sigma},\Theta_{\Delta,\ell}\rangle}{\langle\Theta_{\Delta,\ell},\Theta_{\Delta,\ell}\rangle}\Theta_{\Delta,\ell} \\ 
        k^{\sigma\phi\sigma}b_\phi^2\frac{\langle\mathcal{F}_{0}^{\phi\phi\phi\phi},\Theta_{\Delta,\ell}\rangle}{\langle\Theta_{\Delta,\ell},\Theta_{\Delta,\ell}\rangle}\Theta_{\Delta,\ell} 
        & 
          (1-k^{\phi\sigma\phi})b_\sigma^2\frac{\langle\mathcal{F}_{0}^{\sigma\sigma\sigma\sigma},\Theta_{\Delta,\ell}\rangle}{\langle\Theta_{\Delta,\ell},\Theta_{\Delta,\ell}\rangle}\Theta_{\Delta,\ell} 
      \end{pmatrix} , \\ \\
&
\bm{\mathcal{F}}_\text{susy}
= 
  \sum_{\ell=0}^{\infty}
    \int_{0}^{\infty}
     \frac{ds}{(1-k_\text{susy}^+)(1-k_\text{susy}^-)}
      \begin{pmatrix} 
      b_\sfp^2\frac{\langle\mathcal{F}_{0}^{\sfp\sfp\sfp\sfp},\Upsilon_{\Delta,\ell}\rangle}{\langle\Upsilon_{\Delta,\ell},\Upsilon_{\Delta,\ell}\rangle}\Upsilon_{\Delta,\ell} 
      & 
        -\frac{k_\text{susy}^-\,k_\text{susy}^+}{k^{\sfs\sfp\sfs}}b_\sfs^2\frac{\langle\mathcal{F}_{0}^{\sfsb\sfs\sfs\sfsb},\Upsilon_{\Delta,\ell}\rangle}{\langle\Upsilon_{\Delta,\ell},\Upsilon_{\Delta,\ell}\rangle}\Upsilon_{\Delta,\ell} \\ 
      k^{\sfs\sfp\sfs}b_\sfp^2\frac{\langle\mathcal{F}_{0}^{\sfp\sfp\sfp\sfp},\Upsilon_{\Delta,\ell}\rangle}{\langle\Upsilon_{\Delta,\ell},\Upsilon_{\Delta,\ell}\rangle}\Upsilon_{\Delta,\ell} 
      & 
        (1-k^{\sfp\sfs\sfp})b_\sfs^2\frac{\langle\mathcal{F}_{0}^{\sfsb\sfs\sfs\sfsb},\Upsilon_{\Delta,\ell}\rangle}{\langle\Upsilon_{\Delta,\ell},\Upsilon_{\Delta,\ell}\rangle}\Upsilon_{\Delta,\ell} 
    \end{pmatrix} .          
\end{split}
\end{equation}
Here $\Theta_{\Delta,\ell}$ and $\Upsilon_{\Delta,\ell}$ are conformal and superconformal partial waves, respectively, which are described in \cite{Simmons-Duffin:2017nub} and \cite{Chang:2021fmd}. We also defined the combinations
\begin{equation}\label{eq:kernelevs}
\begin{split}
k^\pm_\text{bos} 
&=
   \frac{1}{2}\, \left( k^{\phi\sigma\phi} \pm \sqrt{\left( k^{\phi\sigma\phi}\right)^2 + 4\,  
  k^{\phi\phi\phi}\, k^{\sigma\phi\sigma}}\right) , \\
k^\pm_\text{susy} 
&=
   \frac{1}{2}\, \left( k^{\sfp\sfs\sfp} \pm \sqrt{\left( k^{\sfp\sfs\sfp}\right)^2 + 4\,  
  k^{\sfp\sfp\sfp}\, k^{\sfs\sfp\sfs}}\right)  . 
\end{split}
\end{equation}
The final step involves computing the inner product of the disconnected correlator with the partial wave. Carrying this out and re-expressing the result in terms of the (super)conformal blocks we arrive at the form quoted in the main text in \eqref{eq:Fcontour}. The bosonic calculation was originally done in \cite{Liu:2018jhs} while the 3d susy calculations can be found in \cite{Chang:2021fmd}.

For completeness let us record the expressions for the kernel eigenvalues. In the bosonic theory we have the eigenvalue matrix
\begin{equation}\label{eq:bosk}
\begin{split}
\begin{pmatrix}
  k^{\phi\sigma\phi}  & k^{\phi\phi\phi} \\
  k^{\sigma\phi\sigma}  & 0 \\
\end{pmatrix}
&=
   \mathfrak{b}_\text{bos}
   \begin{pmatrix}
    \lambda\,\tilde{k}_\text{bos}(\Delta_\phi, \Delta+2\,\Delta_\phi -2\,\Delta_\sigma,\ell) 
    & 
      \frac{b_\phi}{b_\sigma}\,  \lambda\, \tilde{k}_\text{bos}(\Delta_\phi, \Delta+2\,\Delta_\phi -2\,\Delta_\sigma,\ell)  \\
  \frac{b_\sigma}{b_\phi}\, \tilde{k}_\text{bos}(\Delta_\sigma, \Delta+4\,\Delta_\sigma -4\,\Delta_\phi,\ell)  & 0 \\
  \end{pmatrix} ,\\
\tilde{k}_\text{bos}(\Delta_\phi, \Delta,\ell)
&=
  \frac{\pi ^3}{2} \left(1+(-1)^\ell\right) 
   \frac{\Gamma \left(\frac{3}{2}-\Delta _{\phi }\right)^2}{\Gamma \left(\Delta _{\phi }\right)^2}
  \,\frac{\Gamma \left(\frac{3+\ell+\Delta -4 \Delta _{\phi }}{2}  \right)
   \Gamma \left(\frac{-6+\ell-\Delta +8 \Delta _{\phi }}{2}\right)}{ \Gamma \left(\frac{\ell-\Delta +4 \Delta _\phi }{2} \right)\, \Gamma \left(\frac{9+\ell+\Delta -8
   \Delta _{\phi }}{2} \right) } ,
\end{split}
\end{equation}  
while in the susy theory we obtain an analogous expression for the matrix:
\begin{equation}\label{eq:susyk}
\begin{split}
\begin{pmatrix}
  k^{\sfp\sfs\sfp}  & k^{\sfp\sfp\sfp} \\
  k^{\sfs\sfp\sfs}  & 0 \\
\end{pmatrix}
&=
   \mathfrak{b}_\text{bos}
   \begin{pmatrix}
    \lambda\,\tilde{k}_\text{susy}(\Delta_\sfp, \Delta+2\,\Delta_\sfp -2\,\Delta_\sfs,\ell) 
    & 
      \frac{b_\sfp}{b_\sfs}\,  \lambda\, \tilde{k}_\text{susy}(\Delta_\sfp, \Delta+2\,\Delta_\sfp -2\,\Delta_\sfs,\ell)  \\
  \frac{b_\sfs}{b_\sfp}\, \tilde{k}_\text{susy}(\Delta_\sfs, \Delta+4\,\Delta_\sfs -4\,\Delta_\sfp,\ell)  & 0 \\
  \end{pmatrix} ,\\
\tilde{k}_\text{susy}(\Delta_\sfp, \Delta,\ell)
&=
  \pi ^2\, (-1)^{\ell+1} \, 16^{\Delta_\sfp}\, \sin^2(\pi\Delta_\sfp)\, 
  \frac{\Gamma(2-2\Delta_\sfp)^2\, \Gamma \left(\frac{3+\ell+\Delta -4 \Delta _{\sfp }}{2}  \right)
   \Gamma \left(\frac{-4+\ell-\Delta +8 \Delta _{\sfp }}{2}\right)}{ \Gamma \left(\frac{\ell-\Delta +4 \Delta _\sfp }{2} \right)\, \Gamma \left(\frac{7+\ell+\Delta -8
   \Delta _{\sfp }}{2} \right) }\,.
\end{split}
\end{equation}  

Folding in all these results we have the spectral information for the bosonic model:
\begin{equation}
\begin{split}
\mathcal{F}^{\phi\phi\phi\phi} 
&= 
  \sum_{\ell=0}^\infty\, \oint\, \frac{d\Delta}{2\pi i}\, \frac{\rho_\phi(\Delta,\ell)}{1-k_\text{bos}(\Delta,\ell)}\, \mathcal{G}^\text{bos}_{\Delta,\ell} \,,\\ 
\rho_\phi(\Delta,\ell) 
&=
   \left(1+(-1)^\ell\right) 2^{\ell-1}\, \frac{\Gamma(\ell+\frac{3}{2})}{\Gamma(\ell+1)} 
   \,\frac{\Gamma\big(\frac{3}{2}-\Delta_{\phi}\big)^{2}}{\Gamma(\Delta_{\phi})^{2}}\,
     \frac{\Gamma(\Delta-1)\,\Gamma(3-\Delta+\ell)\,\Gamma\big(\frac{\Delta+\ell}{2}\big)^{2}}{\Gamma\big(\Delta-\frac{3}{2}\big)\,\Gamma(\Delta+\ell-1)\,\Gamma\big(\frac{3-\Delta+\ell}{2}\big)^{2}} \, \\
&\qquad\qquad \qquad \qquad 
    \times \frac{\Gamma\big(\Delta_{\phi}-\frac{(\Delta-\ell)}{2}\big)\Gamma\big(\Delta_{\phi}-\frac{3}{2}+\frac{(\Delta+\ell)}{2}\big)}{\Gamma\big(3-\Delta_{\phi}-\frac{(\Delta-\ell)}{2}\big)\Gamma\big(\frac{3}{2}-\Delta_{\phi}+\frac{(\Delta+\ell)}{2}\big)}\,,\\
k_\text{bos}(\Delta,\ell)
&=
    \lambda \, \mathfrak{b}_\text{bos} \, \tilde{k}_\text{bos}(\Delta_\phi, \Delta+6\,\Delta_\phi-6, \ell)
    \left[1+\mathfrak{b}_\text{bos} \, \tilde{k}_\text{bos}(3-2\,\Delta_\phi, \Delta+12-12\,\Delta_\phi, \ell)\right] , \\  
\end{split}
\end{equation}  
with $\mathcal{G}^\text{bos}_{\Delta,\ell}$ being the conformal blocks \cite{Simmons-Duffin:2017nub}. The result for $\mathcal{F}^{\sigma\sigma\sigma\sigma}$ is similar with 
$\rho_\sigma = (1-k^{\phi\sigma\phi}) \, (\frac{1}{\lambda}\rho_\phi|_{\Delta_{\phi} \to \Delta_{\sigma}})$. 
Similarly, for the susy theory we find:
\begin{equation}
\begin{split}
\mathcal{F}^{\sfp\sfp\sfp\sfp} 
&= 
  \sum_{\ell=0}^\infty\, \oint\, \frac{d\Delta}{2\pi i}\, \frac{\rho_\sfp(\Delta,\ell)}{1-k_\text{susy}(\Delta,\ell)}\, \mathcal{G}^\text{susy}_{\Delta,\ell} \,,\\ 
\rho_\sfp(\Delta,\ell) 
&=
    - \frac{2^{4\Delta_\sfp+\ell-1}}{\pi}\, \frac{\sin^2(\pi\Delta_\sfp)}{(\Delta-\ell-1)} 
       \, \sin\left(\frac{\pi}{2}(2\Delta_\sfp+\Delta+\ell)\right) \, \csc\left(\frac{\pi}{2}(\Delta+\ell-2\Delta_\sfp)\right) \, \Gamma(2-2\Delta_\sfp)^2 \\ 
&\qquad
       \times \frac{\Gamma\left(\ell+\frac{3}{2}\right)}{\Gamma(\ell+1)} 
    \frac{\Gamma(\Delta)\,\Gamma(1-\Delta+\ell)}{\Gamma\left(\Delta-\frac{1}{2}\right) \Gamma(\Delta+\ell)} \, \frac{\Gamma\left(\frac{\Delta+\ell}{2}\right)^2}{\Gamma\left(\frac{1-\Delta+\ell}{2}\right)^2} 
    \frac{\Gamma\left(\Delta_\sfp + \frac{\Delta-\ell-2}{2}\right) \Gamma\left(\Delta_\sfp+ \frac{\Delta+\ell-1}{2}\right)}{\Gamma\left(\frac{\Delta-\ell+2}{2} - \Delta_\sfp\right) \Gamma\left(\frac{\Delta+\ell+3}{2}-\Delta_\sfp\right)} \,,
     \\
k_\text{susy}(\Delta,\ell)
&=
    \lambda \, \mathfrak{b}_\text{susy} \, \tilde{k}_\text{susy}(\Delta_\sfp, \Delta+6\,\Delta_\sfp-4, \ell)
    \left[1+\mathfrak{b}_\text{susy} \, \tilde{k}_\text{susy}(2-2\,\Delta_\sfp, \Delta+8-12\,\Delta_\sfp, \ell)\right] . \\  
\end{split}
\end{equation}  
We likewise obtain $\mathcal{F}^{\sfs\sfs\sfs\sfs}$ by the replacement
$\rho_\sfp \to (1-k^{\sfp\sfs\sfp}) \, (\frac{1}{\lambda}\rho_\sfp|_{\Delta_{\sfp} \to \Delta_{\sfs}})$. The off-diagonal correlators can be obtained similarly using the formulae \eqref{eq:4ptfinal}, \eqref{eq:kernelevs}, \eqref{eq:bosk}, and \eqref{eq:susyk}.

As noted in the main text these expressions were used to derive the spectrum. The anomalous dimensions for the bosonic model defined in \eqref{eqn:spectrum} are given at large $\ell$ by
\begin{equation}\label{eq:bosRanomdim}
\begin{split}
\gamma_\phi(\ell,n)
&= 
\frac{16^{\Delta _{\phi }-1}}{\ell^{2 \Delta _{\phi}}}\frac{(-1)^{n+1}}{\pi \Gamma (n+1)} \left(\Delta _{\phi}-2\right) \left(2 \Delta _{\phi }-3\right) \left(2 \Delta _{\phi }-1\right) \left(4 \Delta _{\phi}-3\right) \cos \left(2 \pi  \Delta _{\phi }\right) \\
&\qquad \times\frac{\Gamma \left(2 \Delta _{\phi }-\frac{3}{2}\right)^2  \Gamma \left(3-3 \Delta _{\phi}-n\right)}{\Gamma \left(3-2 \Delta _{\phi}-n\right) \Gamma \left(\Delta _{\phi }-n\right)} \,, \\
\gamma_\sigma(\ell,n)
&= \frac{\Gamma(\Delta_\sigma-n)\Gamma\left(\frac{3}{2}-\frac{3\Delta_\sigma}{2}-n\right)}{\Gamma(3-2\Delta_\sigma-n)\Gamma\left(\frac{3\Delta_\sigma}{2}-\frac{3}{2}-n\right)}\gamma_\phi(\ell,n)\,.
\end{split}
\end{equation}
The anomalous dimensions for the susy model defined in \eqref{eq:susyDeltagam} have nice compact expressions at large $\ell$
\begin{equation}\label{eq:susyRanomdim}
\begin{split}
\gamma_\sfp(\ell,n)
&=
  \frac{(-1)^{\ell}}{\ell^{2-2 \Delta_\sfp }} \, \mathfrak{g}(n,\Delta_\sfp) \,, \\
\gamma_\sfs(\ell,n)
&=
  \frac{2^{1-2 \Delta_\sfs}} {\ell^{2-\Delta_\sfs}}\, 
\frac{\Gamma \left(\Delta_\sfs\right)  \Gamma \left(-n-\frac{3 \Delta_\sfs}{2}+1\right)}{\Gamma\left(-\Delta_\sfs\right) \Gamma \left(-n-\frac{\Delta_\sfs}{2}+1\right)} \, 
\mathfrak{g}(n,\Delta_\sfs)
  \,, \\
\mathfrak{g}(n,\Delta_{_\mathfrak{X}})
&=
  \frac{(-1)^{n}4^{2 \Delta _{_\mathfrak{X} }-1} \left(\Delta _{_\mathfrak{X} }-1\right)  \sin \left(2 \pi  \Delta _{_\mathfrak{X} }\right) \Gamma \left(2-2 \Delta _{_\mathfrak{X} }\right)^2 }{\pi  \Gamma (n+1) \Gamma \left(-n-2 \Delta _{_\mathfrak{X} }+2\right)}\,.
\end{split}
\end{equation}
%

\subsection{Central charges}

The central charges of interest are easily obtained once we know the OPE coefficients. They are given in terms of the OPE data with the conserved currents, the stress tensor for the bosonic theory, and the R-current supermultiplet and flavour supermultiplet for the susy model. 

For example, in the bosonic model the central charge is given by 
\begin{equation}
C_T = \frac{9\Delta_\phi^2}{4\, \abs{C_{\phi\phi T}}^2} \,, \qquad 
\abs{C_{\phi \phi T}}^2 = -\frac{1}{N}\, \Res_{\Delta =3}\left(\frac{\rho_\phi(\Delta,2)}{1-k_\text{bos}(\Delta,2)}\right).
\end{equation}  
Computing the residue we find therefore
\begin{equation}\label{eq:CTbos}
\begin{split}
C_T(\Delta_\phi) 
&= 
  \frac{4N}{5}\, \cot(\pi\Delta_\phi) \bigg[\frac{3}{2\pi} \, \frac{4\Delta_\phi^2(3\Delta_\phi-8) (2\Delta_\phi^2- 5\Delta_\phi+7) + 9(11\Delta_\phi-2)}{(2\Delta_\phi-1)(4\Delta_\phi-3)} \\
& \qquad 
  + 2 \Delta_\phi (\Delta_\phi-2)(\Delta_\phi-1) (2\Delta_\phi-3) \left( (3-2\Delta_\phi) + 2(\Delta_\phi-3) \cos(2\pi\Delta_\phi)\right) \, \csc(4\pi\Delta_\phi)
 \bigg]\,.
 \end{split}
\end{equation}  
Taking the limit $\Delta_\phi \to \frac{1}{2}$ we obtain the result given in \eqref{eq:CTbos0}. On the other hand as $\lambda \to \infty$ (equivalently $\Delta_\phi \to \frac{3}{4}$) we find
\begin{equation}\label{eq:CTnear34}
C_T = 
  N \left[\frac{3}{2} \left(\frac{27}{32}- \frac{87}{80\pi}\right) + \frac{9\sqrt{3}}{256\sqrt{10}\,\pi}
  \left(176-35 \pi - \frac{96}{\pi}\right) \frac{1}{\sqrt{\lambda}} + \order{\lambda^{-1}}
\right] ,
\end{equation}  
which coincides with the value obtained for ${\rm bSYK}^{3d}_{q=4}$ in \cite{Liu:2018jhs} in the strict limit.

Turning to the susy model, the central charge is related to the OPE coefficient for the R-current supermultiplet $\mathcal{R}$, while the flavour central charge is related to the flavour-current multiplet (denoted $\mathcal{J}_f$) OPE coefficient 
\begin{equation}\label{eq:Csusydef}
C_T = \frac{6\,\Delta_\sfp^2}{\abs{C_{\sfpb\sfp\,\mathcal{R}}}^2}\,, \qquad 
C_f = \frac{1}{\abs{C_{\sfpb\sfp\mathcal{J}_f}}^2} .
\end{equation}  
Evaluating the residues we find the explicit expressions:
\begin{equation}\label{eq:Csusy}
\begin{split}
C_T(\Delta_\sfp)
&= 
  -\frac{16N\,\cot(\pi \Delta_\sfp)}{\pi}  \bigg[
 2\pi \, \csc(2\pi \Delta_\sfp) \, \Delta_\sfp \, (\Delta_\sfp-1) \left[
  (\Delta_\sfp-1)\, \sec(2\pi\Delta_\sfp) - \Delta_\sfp +2  \right] +\frac{3\,\Delta_\sfp^2 - 3\Delta_\sfp +1}{1-2\Delta_\sfp} 
  \bigg] ,\\
C_f(\Delta_\sfp) 
&=
  N \bigg[
  4 (\Delta_\sfp -1)\, \sec(2\pi \Delta_\sfp) + \frac{2\cot(\pi \Delta_\sfp)}{\pi (2\Delta_\sfp-1)}
    \bigg] \,. 
\end{split}
\end{equation}
In the limit $\Delta_\sfp \to \frac{1}{2}$ $(\lambda \to 0)$, the central charges are
\begin{equation}\label{eq:Csusy0}
C_T = N\left(6-\frac{32}{\pi^{2}}\lambda+\order{\lambda^{2}}\right) , \qquad C_f = N\left(1-\frac{16}{\pi^{2}}\lambda+\order{\lambda^{2}}\right) ,
\end{equation}
in agreement with the values for $N$ free fields in the strict limit, while in the limit $\Delta_\sfp \to \frac{3}{4}$ ($\lambda \to \infty$) we find
\begin{equation}\label{eq:Csusyinf}
C_T = M\left(6+15\left(\frac{1}{2}-\frac{1}{\pi}\right)\frac{1}{\lambda}+\order{\lambda^{-2}}\right) , \qquad C_f = M\left(4+\left(1-\frac{6}{\pi}\right)\frac{1}{\lambda}+\order{\lambda^{-2}}\right) ,
\end{equation}
which are the values for $M$ free fields in the strict limit (note that the free value for $C_f$ differs from \eqref{eq:Csusy0} by a factor of $4$ because the flavour charges of $\sfs$ and $\sfp$ differ by a factor of $-2$). For $\Delta_\sfp = \frac{2}{3}$, the central charges are given by
\begin{equation}\label{eq:Csusy12}
C_T = N\frac{2^{7}}{3^{2}\sqrt{3}\pi}\left(\frac{2\pi}{\sqrt{3}}-\frac{9}{8}\right) , \qquad C_f = N\left(\frac{8}{3}-\frac{2\sqrt{3}}{\pi}\right) ,
\end{equation}
where $C_T$ is $\frac{3}{2}$ times the value in \cite{Chang:2021fmd}, owing to the fact that the present model has $N+M = \frac{3}{2}N$ chiral multiplets with $R$-charge equal to $\frac{2}{3}$.

The central charges \eqref{eq:Csusy} can be compared with the answer obtained from the supersymmetric localization calculation of the free energy on the squashed sphere $F = -\log Z_{\mathbf{S}_{b}^{3}}$ \cite{Nishioka:2013gza,Gang:2019jut}. The free energy is a function of the IR $R$-charge $\Delta$ of the chiral multiplet and is related to the central charge by
\begin{equation}\label{eq:FtoCT}
C_T^{\text{loc.}}(\Delta) = \frac{48}{\pi^{2}}\Re \frac{\partial^{2} F(\Delta)}{\partial b^{2}}\bigg|_{b=1}\,.
\end{equation}
For a theory of $N$ chiral multiplets with $R$-charge $\Delta_\sfp$ and $M$ chiral multiplets with $R$-charge $\Delta_\sfs$, localization gives the central charge
\begin{equation}\label{eq:CTloc}
C_T^{\text{loc.}} = N\left(C_T^{\text{loc.}}(\Delta_\sfp)+\lambda C_T^{\text{loc.}}(\Delta_\sfs)\right) ,
\end{equation}
which agrees precisely with \eqref{eq:Csusy}. To obtain the flavour central charge, one must turn on a mass deformation $m$ that couples the flavour current to a background gauge field. The flavour central charge is then related to the free energy on the sphere by
\begin{equation}\label{eqn:FtoCJ}
C_f^{\text{loc.}}(\Delta,q) = 8q^{2}\Re \frac{\partial^{2} F(\Delta)}{\partial m^{2}}\bigg|_{m=0,b=1}\,,
\end{equation}
which depends on the IR $R$-charge $\Delta$ and flavour charge $q$, leading to flavour central charge
\begin{equation}\label{eq:CTloc}
C_f^{\text{loc.}} = N\left(C_f^{\text{loc.}}(\Delta_\sfp,1)+\lambda \,C_f^{\text{loc.}}(\Delta_\sfs,-2)\right) ,
\end{equation}
agreeing again with \eqref{eq:Csusy}.

\end{widetext}
\end{document}

%% file: 3dvector-refs.tex
%